# Excitonic lasing in semiconductor quantum wires

L. Sirigu, D. Y. Oberli, L. Degiorgi, A. Rudra, and E. Kapon

Department of Physics, Swiss Federal Institute of Technology-EPFL, CH-1015 Lausanne, Switzerland

Direct experimental evidences for excitonic lasing is obtained in optically pumped V-groove quantum wire structures. We demonstrate that laser emission at a temperature of 10 K arises from a population inversion of localized excitons within the inhomogenously-broadened luminescence line. At the lasing threshold, we estimate a maximum exciton density of about $1.8 \times 10^5 cm^{-1}$.

Semiconductor lasers incorporating low-dimensional heterostructures, quantum wires and quantum boxes, are attracting considerable interest because of their potential for improved performance over quantum well lasers[1]. This prediction is based, in the single-particle picture, on the sharper density of states resulting from the confinement of the charge carriers in two or three directions. The inclusion of electron-hole Coulomb correlations in the theoretical description of the optical spectra of semiconductor quantum wires has, however, dampened these early expectations. The primary effect of Coulomb correlations is to remove the singularity of the one-dimensional (1D) joint density of states and to greatly reduce the absorption above the band edge[2]. In the regime of high carrier densities screening of the Coulomb interaction, band gap renormalization and phase space filling might become important[3]. The evolution of the optical response as the density of electron-hole pairs is increased has been intensively studied in 1D systems in recent years. However, contrasting theoretical predictions were made: in several studies[4-5], gain was found to appear only at densities above the Mott density (about mid $10^5$ cm$^{-1}$) while, in another one, substantial gain was calculated at densities largely below this value when a self-consistent treatment of the electron-hole correlations was included[6]. The occurrence of excitonic gain that is predicted by the latter study is an interesting feature of 1D quantum structures. Laser emission from excitons confined in quantum wells has so far not been observed in III-V semiconductor heterostructures, but has been observed in their II-VI counterparts for which the exciton binding energy is much larger[7]. Evidence for lasing attributed to excitons has been previously reported in T-shaped quantum wire structures (QWR's), for which only one electron 1D-subband was present[8].

In the present study, we report on the observation of lasing from excitons in optically excited V-groove GaAs/AlGaAs QWR laser structures. The emission is attributed to the recombination of excitons associated with the lowest energy electron- and hole- subbands of the QWR. Moreover, we find that the emission energy remains nearly constant within the inhomogenously broadened photoluminescence line of the QWRs for both continuous wave (cw) and pulsed optical excitation over a wide range of power densities. These results corroborate the important role played by electron-hole Coulomb correlations in the optical emission from quasi-1D quantum wires in the density regime of the Mott transition.

The sample used for our study was grown by low-pressure organometallic chemical vapor deposition on a GaAs substrate patterned with a 3 μm-pitch grating. It is a semiconductor laser structure that incorporates five vertically stacked GaAs quantum wires separated by 47 nm thick $Al_{0.25}Ga_{0.75}As$ barriers placed in the core region of an optical wave guide. The core region has a total thickness of 370 nm and is cladded with 1 µm thick $Al_{0.62}Ga_{0.38}As$ layers on both sides. The center of the QWRs-stack was displaced by 72 nm from the center of the core region in order to obtain an optimal mode confinement factor[9]. The position of the QWRs-stack can be seen in the cross-sectional transmission electron micrograph of the core region of our QWR laser structure depicted in Fig.1. The V-shape of the dielectric waveguide yields an optical mode in the shape of a heart that is well confined in the core region of the waveguide according to our calculation of the electromagnetic field distribution of the optical modes[9].

Cleaved optical cavities were mounted in a helium-flow cryostat and kept at 10 K. The laser cavities were optically pumped using 3 ps pulses from a mode-locked Ti-Sapphire laser operating at a wavelength of 720 nm (1.72eV) with a repetition rate of 80 MHz. The excitation light was incident on the growth surface and focused through two cylindrical lenses onto a stripe 46 µm wide and 1.5 mm long oriented in the direction parallel to the optical wave guide. The QWRs were perpendicular to the cleaved facets forming the mirrors of the optical cavity. The emitted light was collected from one of the cleaved facets, dispersed by a double-grating spectrometer and detected with a cooled GaAs photomultiplier tube.

Optical emission spectra of the QWR laser structure are displayed in Fig. 2 for different values of the optical power density below, at and above the threshold for lasing in the QWR. Upon increasing the pump power, we observe a nearly constant energy of the peak at 1.581 eV that corresponds to the optical transition $e_1$-$h_1$ associated with the ground electron- and hole-subband of the QWRs. A significant spectral narrowing is also found as the power density is increased and crosses the lasing threshold. This evidences the existence of amplified spontaneous emission within this inhomogenously broadened PL line in this density regime. A typical light output versus input power characteristic is shown in Fig. 2 b for a cavity length of about 0.9 mm. The emission intensity varies linearly at low excitation power over three orders of magnitude (from 0.1 to 100 mW). Above the lasing threshold (at 350 mW) the intensity variation is again linear, indicating that the modal gain has saturated. In Fig. 2 c, a high-resolution emission spectrum obtained above threshold features well-resolved Fabry-Perot modes that correspond to different longitudinal optical modes of the cavity within the inhomogenuous line of the QWR-PL. It is worth noting that the optical oscillations have in this spectrum a high contrast ratio, the value of which being mainly limited, however, by the non-uniform pumping profile of the focussed laser beam.

In order to identify the regime in which lasing takes place, we now estimate the electron-hole pair density that was directly generated in each QWR by the 3 ps optical pulse. The electron-hole pair density is given by

$$n = \alpha t(1-R)(Pw/\hbar\omega)T(\Gamma_{ho}/\Gamma_{inho})$$

where $\alpha$ is the excitonic absorption coefficient, t and w are respectively the effective wire thickness and width, R (0.3) is the normal incidence reflectivity at the laser energy $\hbar\omega$, P is the power density, T is the pulse repetition period (12.5 ns). The last term in the previous expression is the ratio between the homogeneous ($\Gamma_{ho}$) to inhomogeneous ($\Gamma_{inho}$) widths of the optical transition. Assuming an absorption coefficient, $\alpha$, of 16000 cm$^{-1}$ in the V-groove QWR and a power density, P, of 0.6 kW/cm$^2$ at the lasing threshold, we estimate an excited electron-hole pair density of 1.8 10$^5$ cm$^{-1}$.[10-11] This density represents an *upper bound* of the inverted population achieved for lasing. Because this density is lower than the Mott transition density we conclude that the optical emission is dominated by excitonic recombination. We emphasize that actual theoretical estimates[3,4] of the Mott density yield different values between 3 and 8 x10$^5$ cm$^{-1}$ for GaAs QWR structures with the same exciton binding energy. To further confirm the excitonic nature of the emission we compare, in Fig. 3, PL spectra that are obtained under pulsed excitation over a wide range of power densities with a PL spectrum obtained under cw excitation. In this latter case the power density is very low and corresponds to about one tenth the typical power density used in the PL and PL-excitation (PLE) spectra displayed in Fig. 4. The relevance of excitonic recombination at low temperature in the PL and PLE spectra has already been established in previous studies of similar QWR structures[12]. *From the absence of any significant shift (less than 2 meV) we infer that the lasing emission originates from the recombination of excitons as it is the case for the QWR-peak of the cw-PL spectrum.*

We will now address the origin of the energy blueshift that is observed as the power density is increased up to about 100 W/cm$^2$ (below lasing threshold). It could originate from an imperfect compensation between the renormalization of the band-gap energy introducing a red-shift (BGR) and a lowering of the exciton binding energy yielding a blue-shift as the carrier density is increased. In the density range of the present experiment, these many-body effects were evaluated resulting in an absorption peak that exhibits a weak blueshift as the carrier density approaches the Mott density[3]. Alternatively, the energy blueshift may be attributed to effects related to disorder. We note that, under cw excitation condition, similar shifts were indeed observed along with the appearance of a high energy tail in the PL spectrum that is indicative of a heating of the hot carrier distribution[14]. These latter findings can also be related to disorder effects that broaden the peaked optical density, thereby yielding a blueshift of the emission for a heated Maxwell-Boltzmann distribution of electron-hole pairs[15]. Beyond the lasing threshold an energy redshift is finally observed, which is caused by a local heating of the sample.

The PL and PLE spectra, which are depicted in Fig. 4, were obtained from the same sample in a standard pseudo-backscattering configuration at a temperature of 10 K[16]. In this sample, the upper part of the optical wave guide was chemically etched away including the top and part of the side quantum wells in order to strongly reduce their contributions to the PL spectrum. The energy position of the PL peak was again found at 1.582 eV. The PLE spectrum shows six prominent peaks corresponding to optical transitions between electron and hole 1D-subbands of the same index. The optical spectra are characterized by a full width at half maximum of 9 meV (lowest optical transition) and a Stokes shift of 6 meV. The comparison of the PL and PLE spectra demonstrates unambiguously that the lasing emission at 1.581 eV correspond to the $e_1$-$h_1$ transition between the respective n=1 subband of electrons and holes. The presence of a significant Stokes shift indicates that, at the energy of the lasing peak, the excitons are localized. Localization is induced by the presence of interfacial disorder in these QWRs. Disorder can strongly affect the optical properties of the QWRs : previous studies performed on similar V-groove QWRs have shown that the radiative lifetime of localized excitons is enhanced over that of free excitons[17-18] and that it is more weakly dependent on temperature than the predicted dependence for free excitons[15]. Thus, we infer that the observed laser emission arises from a population inversion of localized excitons on the low-energy side of the absorption peak.

Excitonic gain in an inhomogenous emission line can arise if the exciton oscillator strength is large enough: otherwise, as the carrier density is increased, significant bleaching of the excitonic resonance might occur before a large enough population inversion is realized in order to reach the gain regime. It is also worth noting that the build-up of a population inversion from localized excitons does not need to rely on an effective spectral diffusion to the relevant localized states within the inhomogenous luminescence line. However, the relaxation into these localized sites must occur on a time scale that is shorter than the average radiative lifetime of the excitonic population. Further experimental work and theoretical development are deemed necessary to address the dynamical aspect of the gain generation.

In conclusion, we have demonstrated laser emission that is sustained by excitonic gain at a temperature of 10 K in a semiconductor quantum wire structure. Laser emission is shown to arise from the population inversion of localized excitons. Whether disorder and hence localization of the excitons is necessary to obtain excitonic gain remains to be explored as our understanding of the gain generation in a weakly-disorded semiconductor gain medium is not yet complete. This study should stimulate further a theoretical description of the effects of electron-hole Coulomb correlations in the gain regime of quasi one-dimensional semiconductor systems.

We wish to acknowledge helpful discussions on the excitonic gain in 1D systems with C. Piermarocchi. We are also grateful to K. Leifer for providing the TEM micrographs. This work was supported in part by the Swiss National Foundation for Sciences.

Figures

Figure 1: Cross-sectional dark-field micrograph of the core region of the QWR laser structure obtained by transmission electron microscopy. Brighter areas correspond to regions with a larger Al composition. The location of the five QWRs is marked by an arrow.

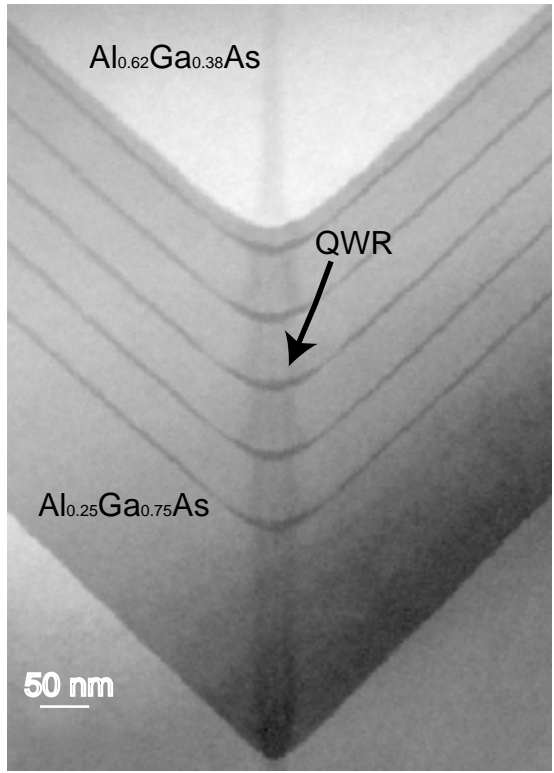

Figure 2: (a) Photoluminescence spectra at 10K of the QWR laser sample above, below and near the lasing threshold in TE-polarization. (b) Dependence on input excitation power of the PL output power; arrows indicate the excitation powers used for the optical spectra depicted in (a). (c) High-resolution emission spectrum above the lasing threshold showing the Fabry-Perot modes of the optical cavity (the mode linewidth is equal to 0.37 Å and the instrumental resolution is set to 0.17 Å).

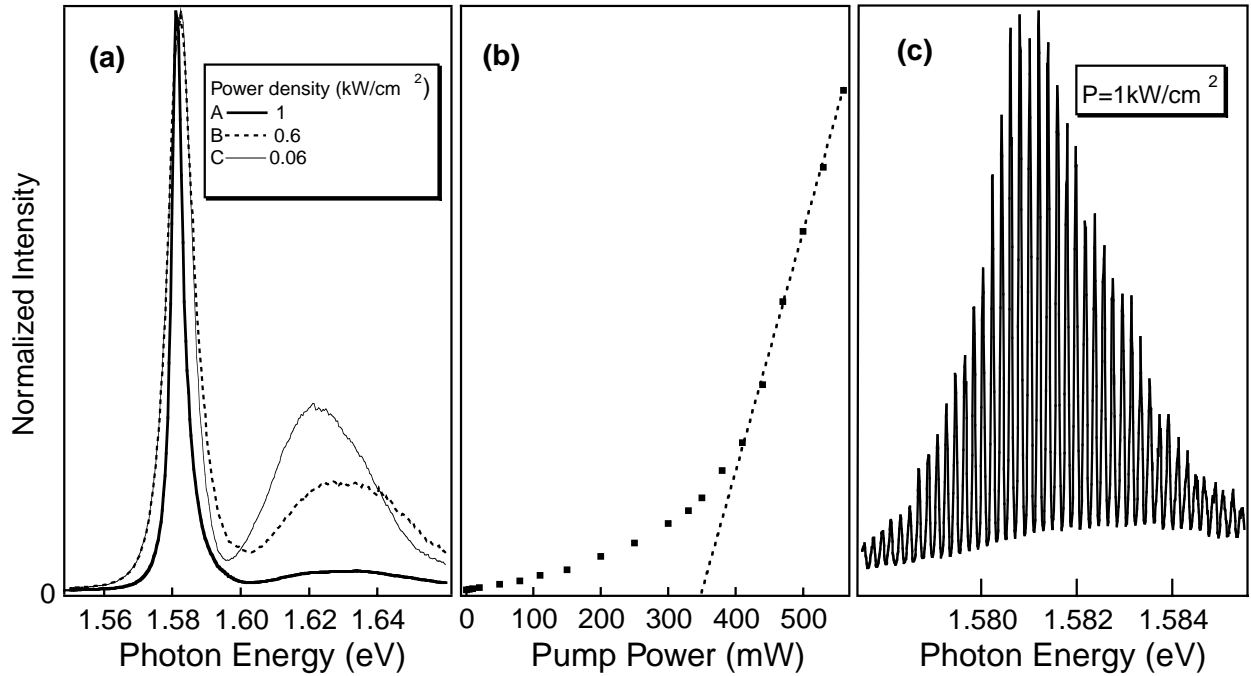

Figure 3 : Evolution of the photoluminescence spectra taken under pulsed excitation with increasing power densities and cw-PL spectrum obtained from the cleaved edge at the same position. The energy shift of the PL emission remains below the value of the Stokes shift (6 meV). Spectra have been shifted upwards for clarity. Estimated exciton density under cw excitation is about 10 cm$^{-1}$.

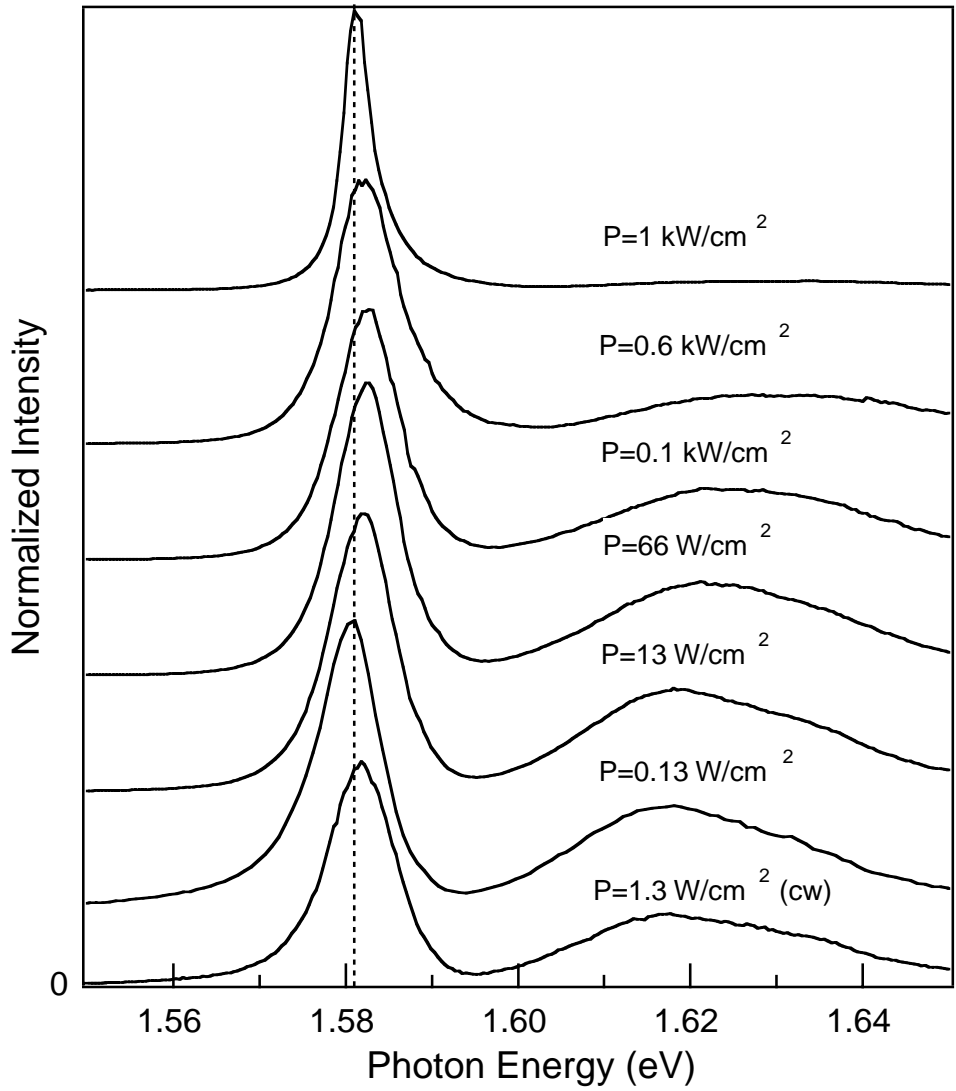

Figure 4: Linearly-polarized PLE spectrum and the corresponding PL spectrum of an etched QWR laser sample at 10K. The polarization of the excitation is parallel to the wire axis. The different optical transitions $e_n$-$h_n$ are marked by arrows.

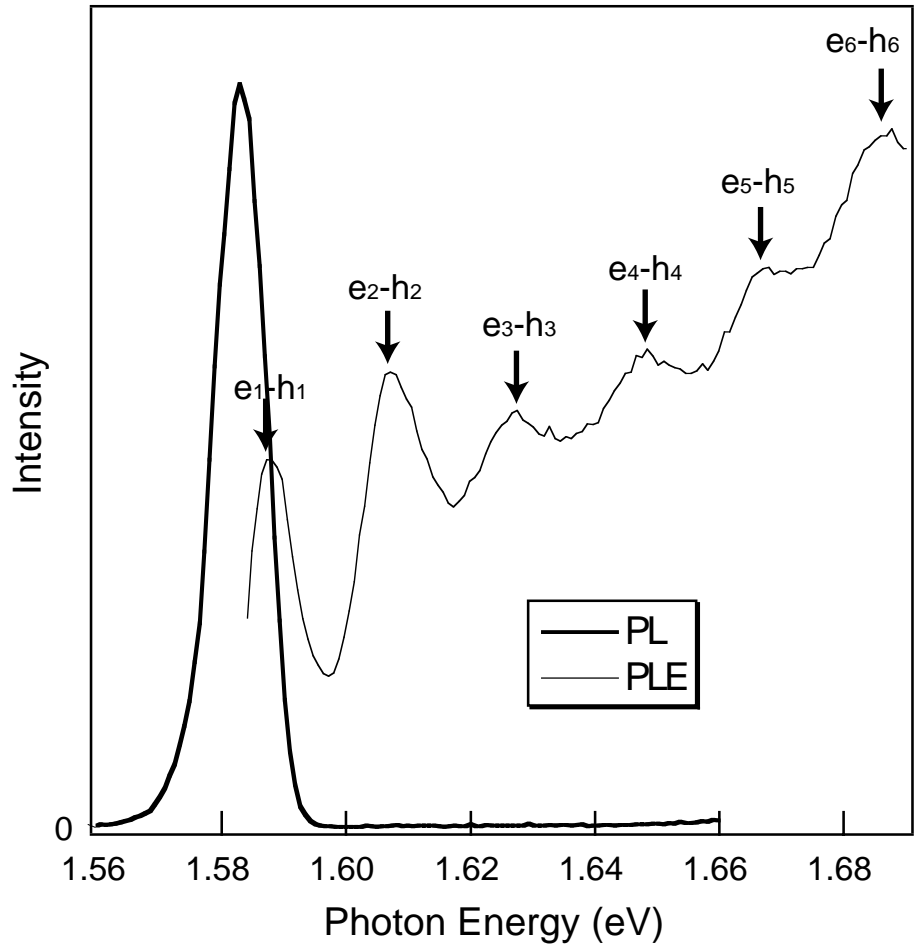